\documentclass[aps,prd,superscriptaddress,showpacs,amsmath,amssymb]{revtex4}
\usepackage{graphicx,epsf}
\usepackage{amsfonts}
\usepackage{amssymb}
\usepackage[]{latexsym}
\newcommand{\be}{\begin{eqnarray}}
\newcommand{\ee}{\end{eqnarray}}

\newcommand{\ket}[1]{\mbox{$\mid #1\,\rangle$}}

\newcommand{\nRband}{n_{Rband}}
\newcommand{\nIP}{n_{IP}}
\newcommand{\lamosc}{\lambda_{4D}}
\newcommand{\Nlam}{N_\lambda^{(4D)}}
\newcommand{\lamoscFiveD}{\lambda_{5D}}
\newcommand{\NlamFiveD}{N_\lambda^{(5D)}}

\newcommand{\umax}{|u_{\rm max}|}
\newcommand{\eps}{\epsilon}

\newcommand{\srarr}{\rightarrow}
\newcommand{\duzero}{\dot{u}_0}
\newcommand{\dxzero}{\dot{x}_0}
\newcommand{\Eres}{E_{\rm Res}}
\newcommand{\ELres}{(LE)_{\rm Res}}
\newcommand{\beq}{\begin{equation}}
\newcommand{\eeq}{\end{equation}}
\newcommand{\bu}{\duzero^{\max}}

\begin{document}

\begin{flushright}
DO-TH-09/08
\end{flushright}

\title{
Baseline-dependent neutrino oscillations with extra-dimensional shortcuts
}

\author{Sebastian Hollenberg}
\email{sebastian.hollenberg@uni-dortmund.de}
\affiliation{Fakult\"at f\"ur Physik, Technische Universit\"at Dortmund, D-44221 Dortmund, Germany}

\author{Octavian Micu}
\email{octavian.micu@tu-dortmund.de}
\affiliation{Fakult\"at f\"ur Physik, Technische Universit\"at Dortmund, D-44221 Dortmund, Germany}

\author{Heinrich P\"as}
\email{heinrich.paes@uni-dortmund.de}
\affiliation{Fakult\"at f\"ur Physik, Technische Universit\"at Dortmund, D-44221 Dortmund, Germany}

\author{Thomas J. Weiler}
\email{t.weiler@vanderbilt.edu}
\affiliation{Department of Physics and Astronomy, Vanderbilt University, Nashville,
Tennessee 37235, USA}

\begin{abstract}
In extra-dimensional scenarios oscillations between active and sterile neutrinos can be governed by a new resonance
in the oscillation amplitude.
This resonance results when cancelation occurs between two phase differences, the usual kinematic one coming
from the neutrino mass-squared difference, and a geometric one coming from the
difference in travel times of the sterile neutrino through the bulk relative to the active neutrino confined to the brane.
In this work we introduce a specific metric for the brane-bulk system,
from which we explicitly derive extra-dimensional geodesics for the sterile neutrino,
and ultimately the oscillation probability of the active-sterile two-state system.
We find that for an asymmetrically-warped metric,
the resonance condition involves both the neutrino energy $E$ and the travel distance $L$ on the brane.
In other words, the resonant energy may be viewed as baseline-dependent.
We show that to a good approximation, the resonance condition is not on $E$ or on $L$,
but rather on the product $LE$.
The model is rich in implications, including the possibility of multiple solutions to the resonance condition,
with ramifications  for existing data sets, e.g., LSND and MiniBooNE.
Some phenomenology with these brane-bulk resonances is discussed.
\end{abstract}
%
%
\maketitle

\section{Introduction}

In theories with large extra dimensions, the Standard Model particles are typically confined
to the $3+1$-dimensional brane,
which is embedded in an extra-dimensional bulk \cite{ArkaniHamed:1998rs, Randall:1999ee}.
Singlets under the gauge group however are allowed to travel freely in the bulk
as well as on the brane.
\par
It has been argued that the LSND neutrino oscillation anomaly \cite{Aguilar:2001ty} and
the MiniBooNE~\cite{AguilarArevalo:2007it} null result~\footnote{The model described in this work relies on metric shortcuts, which we will explain in detail in the following sections, identical for singlet particles and their antiparticles. Thus, the model predictions are CP-symmetric. In particular, the model therefore cannot accommodate the recent MiniBooNE claims that an excess of flavor changing events exists in the neutrino channel~\cite{AguilarArevalo:2007it} but not in the antineutrino channel~\cite{AguilarArevalo:2009xn}.} 
might be explained by a brane-bulk
resonance in active-sterile neutrino oscillations~\cite{Pas:2005rb}.
The resonance arises due to the additional phase difference $\delta (Ht)=t \delta H+ H\delta t$
induced when the sterile neutrinos
take temporal shortcuts through an extra dimension.
In general, $H$ is an $n\times n$ matrix for $n$ neutrino flavors.
\par
Thus, in these models there are two sources of phase difference,
the standard one $t\delta H= L\Delta m^2/2E$, and a new one $Ht\,(\delta t/t)$ arising from
temporal shortcuts through the bulk available to gauge singlet quanta.
The two phase differences may beat against one another to produce resonant phenomena.
Here, as in~\cite{Pas:2005rb}, we will focus on the gauge singlet sterile neutrino~\cite{Cirelli:2004cz}.
\par
For simplicity, the shortcut parameter $\eps\equiv \delta t/t$ in~\cite{Pas:2005rb} was taken to be a constant.
It was given a physical interpretation as the ``aspect ratio''
of a typical brane fluctuation; quantitatively, it was related to the fluctuation geometry as
$\eps =\left( \frac{\pi}{2}\frac{\delta u}{\delta x} \right)^2$,
where $\delta u$ is the amplitude of the brane fluctuation in the bulk direction $u$,
and $\delta x$ is the length of the fluctuation along the brane direction $x$.
It is the purpose of this article to extend the idea in~\cite{Pas:2005rb} to an asymmetrically-warped metric framework for the
extra dimension.
Significant new physics emerges from this generalization.
It will be seen that the shortcut parameter
is no longer a constant, but rather depends on the neutrino propagation distance,
i.e., on the experimental baseline length $L$.
Larger baselines allow more effective shortcuts
(larger $\eps$), which leads to
correspondingly smaller resonance energies.
This effect can be traced back to the travel times of the neutrinos. 
Longer neutrino travel times associated with longer baselines on the brane
allow the off-brane geodesic of the sterile neutrino to plunge deeper into the bulk and experience a greater 
warp factor.
The smaller resonance energies, in turn, suppress active-sterile neutrino mixing at large energies.
Thus, one expects  little or no active-sterile oscillations at the relatively large baselines
accompanying atmospheric and astrophysical neutrino sources.
\par
For the asymmetrically-warped metric which we investigate, a remarkable feature emerges.
It turns out that to a good approximation,
the location of the resonance depends on the {\sl product} of baseline and energy, $LE=\ELres$,
where $\ELres$ is a constant involving the warp factor and neutrino mass and mixing angle.
From this result one infers that
experiments with $LE$ well below $\ELres$ should observe a 5D version of active-sterile vacuum oscillations
(presented in section~\ref{subsec:5Dlimit}),
while experiments with $LE$ well above $\ELres$ should have active-sterile mixing suppressed
below observable levels.
\par
Another feature of the new phenomenology presented herein,
is the possibility of multiple brane-bulk resonances, with interference among them.
This is because there are multiple extra-dimensional paths contributing to the
propagation amplitude of the sterile neutrino, each with a different intrinsic resonant energy.
The result of a quantum mechanical sum of paths is a richer picture for the resonant behavior than
that of the sharp resonant energy in~\cite{Pas:2005rb}, or than that which occurs in neutrino propagation
in a nonzero matter density \cite{Wolfenstein:1977ue, Mikheev:1986wj}.
\par
In the next section we review the physics of the brane-bulk resonance,
to set the stage for the further developments presented in this work.

\section{Two-state active-sterile oscillations in a brane-bulk scenario}\label{sec:twostate}

Following \cite{Pas:2005rb}, we consider two mass eigenstates $\nu_1$ and $\nu_2$,
and a small active-sterile mixing angle $\theta$.
The active-sterile mixing angle $\theta$ relates the
flavor eigenstates $\nu_a$, $\nu_s$ with the mass eigenstates
$\nu_1$, $\nu_2$ via a unitary transformation
    \be
        \ket{\nu_{\alpha}} = \sum\limits_{i=1}^2 U^{\ast}_{\alpha i} \ket{\nu_i} \qquad
        \text{with} \quad \alpha = \text{a, s}
    \ee
and
    \be
        U = \begin{pmatrix} \ \ \ \cos\theta & \sin\theta \\ -\sin\theta & \cos\theta \end{pmatrix}.
    \ee
With small mixing angle,
the state $\nu_a$ is mostly $\nu_1$, and the state $\nu_s$ is mostly $\nu_2$.
\par
The Schr\"odinger equation in flavor space describing
two-state neutrino oscillations reads
    \be
        \frac{d}{dt} \left(\begin{array}{c}\nu_a(t) \\ \nu_s(t)\end{array}\right) = H \ \left(\begin{array}{c}\nu_a(t) \\ \nu_s(t)\end{array}\right)\,,
    \ee
where the relevant Hamiltonian is given by
    \be
        H'=\frac{\Delta m^2}{4E} \begin{pmatrix} -\cos2\theta & \sin2\theta \\
        \ \ \ \sin2\theta & \cos2\theta \end{pmatrix} -\frac{1}{2E} \begin{pmatrix} 0 & 0 \\
        0 & B \end{pmatrix}\,.
    \ee
The ``induced mass-squared difference" is
    \be
        B(E)=2E^2\epsilon\,,
    \ee
and we have dropped the irrelevant piece of $H$ equal to
$E + (m_2^2+m_1^2)/4E$ times the identity matrix.
The parameter $\epsilon\equiv \delta t/t$ is the shortcut parameter,
which parameterizes the relative difference in travel times of
active and sterile neutrino flavors.
It  will be derived explicitly for an asymmetrically-warped metric in section~\ref{sec:metric}.
The novel feature of this work will be that $\eps$, and the resonant energy $\Eres$ that results from it,
will show dependencies on the baseline length $L$.
\par
This Hamiltonian system bears much resemblance to the
Mikheev-Smirnov-Wolfenstein~(MSW) Hamiltonian~\cite{Wolfenstein:1977ue, Mikheev:1986wj}
describing neutrino oscillations in matter.
Accordingly, we shall henceforth
refer to $\theta$ and $\Delta m^2$ as the 4D~vacuum mixing angle
and 4D~vacuum mass-squared difference, respectively.
In the brane-bulk system, the shortcut in the extra dimension has been parameterized by an
effective potential with nonzero sterile-sterile component in the flavor space Hamiltonian.
This term is the analog of the $\nu_e$-$\nu_e$ potential term  in the MSW Hamiltonian,
induced by coherent elastic forward scattering of neutrinos on electrons.
\par
Note, however, that there are three important
differences between the brane-bulk system and the matter system.
First of all, as previously mentioned, gravitationally determined geodesics do not
discriminate between particle and antiparticle and the effective
potential is therefore the same for neutrino and antineutrino.
Secondly, the energy dependence is stronger for the brane-bulk system than for the matter system.
In the brane-bulk system, the induced mass-squared difference varies as  $E^2$,
whereas in the MSW system~\cite{Mikheev:1986wj} the variation is proportional to $E$.
The third difference is that in the brane-bulk system, there is no time or space dependence in the
effective potential, and therefore none in the Hamiltonian.
\par
Diagonalization of the  $2\times2$  effective Hamiltonian proceeds analogously to the MSW system.
One finds a new mixing angle $\tilde\theta$, expressible in terms of
4D~vacuum values as
    \be\label{tan2tilde}
        \tan2\tilde\theta = \frac{\tan2\theta}{1-\frac{E^2}{\Eres^2} }\,
    \ee
 with
    \be
        \Eres = \sqrt{\frac{\Delta m^2 \cos2\theta}{2\epsilon}}.\label{Eres}
    \ee
As with MSW, neutrino mixing can be maximal ($\tilde{\theta}=\pi/4$)
even for a small 4D~vacuum mixing angle.
\par
Eq.~(\ref{Eres}) can be rewritten as
		\be\label{smalleps}
				\epsilon=\frac{\cos 2\theta}{2}\frac{\Delta m^2}{\Eres^2}\,,
		\ee
which shows that $\epsilon$ will be very small for $\Eres^2 \gg \Delta m^2$.
In~\cite{Pas:2005rb}, $\Delta m^2$ was taken to be $\sim {\rm eV}^2$ to accommodate the
LSND data, and $\Eres$ was taken above 30~MeV and up to the
MiniBooNE excess region of $\sim 300$~MeV.
Consequently, $\eps$ turned out to be about  $0.5\times 10^{-16\pm 1}$.
\par
The existence of a resonant energy naturally
divides the energy domain into three regions. Above the resonance
at $E \gg E_{\text{Res}}$, the sterile state decouples from the active
state as $\sin^2 2\tilde\theta \to 0$, and therefore oscillations
are suppressed in this regime.  At resonance, i.e for $E=E_{\text{Res}}$,
the mixing angle $\tilde\theta$ attains a maximum.
Finally, below the resonant energy, the oscillation parameters reduce to their
4D~vacuum values.
\par
The new eigenvalue difference $\delta\tilde H$ is given by
    \be
        \delta\tilde H = \frac{\Delta m^2}{2E}\sqrt{\sin^22\theta + \cos^2 2\theta
        \left[1-\frac{E^2}{\Eres^2}\right]^2} \label{tildeH}
    \ee
and one obtains the usual expression for the
flavor-oscillation probability
    \be
        P_{as} = \sin^2 2\tilde\theta \ \sin^2 \frac{L\delta\tilde{H}}{2} \label{pas}
    \ee
with the expression
    \be
        \sin^2 2\tilde\theta = \frac{\sin^2 2\theta}{\sin^22\theta + \cos^2 2\theta
        \left[1-\frac{E^2}{\Eres^2}\right]^2} \label{tildesin}
    \ee
following from Eq.~(\ref{tan2tilde}).
\par
Such is the picture for a sterile neutrino traversing a unique shortcut through the bulk.
In the next section, we avoid the nebulous feature of a ``typical fluctuation'' of the brane in the bulk by
introducing a metric for the brane-bulk system.
This will allow us to solve explicitly for the brane-bulk geodesics (plural) of the sterile neutrino.
We will find that there are multiple geodesic paths available to the sterile neutrino.
Each path corresponds to an amplitude, and
the multiple paths necessitate a quantum mechanical sum,
which complicates but enriches the analysis.

\section{A metric and geodesics  for sterile neutrino shortcuts}
\label{sec:metric}

A sterile neutrino which propagates in
the bulk as well as on the brane will in general have a different trajectory
compared to the active flavor which is confined to the brane.
If the extra dimension is ``warped'' relative to the brane dimensions, then
the geodesic of the sterile neutrino will be shorter than the geodesic of the active neutrino.
For an observer on the brane, it will appear as if the sterile neutrino has taken
a shortcut through the extra dimension.
Such apparent superluminal behavior for gauge-singlet quanta has been discussed earlier
for the graviton~\cite{Kaelbermann:1999jw, Chung:1999xg}.
\par
Specifically, we consider an asymmetrically warped $4+1$-dimensional spacetime
metric of the form
     \begin{eqnarray}
         d\tau^2 = dt^2 - \sum_{i=1}^{3} \eta^{2}(u) \left(dx^i\right)^2 - du^2 \label{metric},
     \end{eqnarray}
where $u$ is the extra dimension and $\eta(u)$ is the warp factor.
Since space but not time is warped, the warp is termed ``asymmetric''~\cite{Chung:1999xg, Csaki:2000dm}.
The Standard Model neutrinos live on the brane while the sterile
neutrinos propagate freely in the extra dimension. Our observable
brane is the $u = 0$ submanifold, which we take to be globally Minkowskian, i.e. $\eta(u=0)=1$.
We choose a warp factor of the form
$\eta(u) = e^{-k\left|u\right|}$, with $k$ assumed to be a (presently) unknown constant of Nature
with dimension of inverse length~\footnote
{
As written, the metric element $g_{xx}=e^{-2k |u|}$ is non-differentiable at $u=0$.
This may be cured by smoothing $g_{xx}$ around the point $u=0$.
Such smoothing does not affect our results.
The important feature of $g_{xx}$ for our work
is the $u \leftrightarrow -u$~symmetry.
}.
Standard Model neutrinos live in the 4D~Minkowski spacetime, while the
sterile neutrinos experience the full five dimensional metric.
We may choose the direction of the brane component of neutrino velocity to be along $x$;
this allows us to set $dy$ and $dz$ to zero from here on.
Our line element is reduced to
    \be\label{line-element}
        d\tau^2 = dt^2 - e^{-2k |u|} \, dx^2 - du^2\,.
    \ee
We choose the absolute value of $u$ in
the warp factor such that the bulk acts like a converging lens on both
sides of the brane, returning the sterile neutrino to the brane where it may interfere with the
active neutrino.
\par
Our task is to calculate the difference $\delta t =t\,\eps$ in the
propagation time between the Standard Model (active) neutrinos which
propagate only along our brane, and the sterile neutrinos which take
shortcut excursions through the warped extra dimension.
For the active neutrinos, we have immediately the Minkowski space result
$t^{\text{brane}}=L/\beta$, where $\beta$ is the speed of the neutrino.
For the sterile neutrinos, we must find and solve geodesic equations.
It is well-known that geodesic equations may be found by
treating $d\tau^2$, or equivalently, $(d\tau/d\lambda)^2$, with $\lambda$ any
affine parameter which marks progress along the geodesic, as a Lagrangian subject to the
extremization conditions given by the Euler-Lagrange equation.
Here we have
    \be\label{Lagrangian}
        L= \left( \frac{dt}{d\lambda} \right)^2 - e^{-2k|u|} \left( \frac{dx}{d\lambda} \right)^2 -
   	    \left( \frac{du}{d\lambda} \right)^2\,.
    \ee
There result three Euler-Lagrange/geodesic equations, one for each of the variables $\{t, x, u\}$.
They are the following set of coupled nonlinear differential equations
    \be
        \frac{d^2t}{d\lambda^2}&=&0 \label{geo1}\\
        \frac{d}{d\lambda} \left( e^{-2k|u|} \frac{dx}{d\lambda} \right) &=& 0 \label{geo2}\\
        \frac{d^2u}{d\lambda^2} \pm ke^{-2k|u|}\left(\frac{dx}{d\lambda}\right)^2&=&0 \label{geo3}.
    \ee
The $\pm$ signs refer to trajectories in the positive or negative half-planes of $u$.
By integrating Eq. (\ref{geo1}) twice, one readily infers that
$\lambda$ and $t$ are proportional to one another.
We call the proportionality constant $dt/d\lambda\equiv\gamma$.
It is most easily obtained from the line element in Eq.~(\ref{line-element}), by taking $d\tau$ to be $d\lambda$,
as is appropriate for massive particles. One gets
    \be\label{gamma}
        \gamma^{-2}=\left( \frac{d\tau}{dt} \right)^2 = 1-e^{-2k |u|}\,\dot{x}^2 - \dot{u}^2\,,
    \ee
and we have denoted some occurrences of $d/dt$ by an overdot, to streamline notation.
Initially, the neutrinos are created on the brane, where $u=0$, so we have
    \be
        \gamma^2=\frac{1}{1-\beta^2},\quad {\rm with\ } \beta=\sqrt{\dot{x}^2_0 +\dot{u}^2_0}\,.
    \ee
A variable with a zero subscript denotes an initial ($t=0$) value.
The interpretation of $\gamma$ and $\beta$ is now clear:
$\beta$ is the neutrino speed, and $\gamma$ is the usual boost factor, equal to $E/m$.
The only difference for the sterile neutrino versus the active neutrino is that the sterile is allowed to
have an initial velocity component $\dot{u}_0$ in the bulk direction, while the active neutrino cannot.
\par
Translational invariance is maintained on the brane
and in fact, the Minkowski metric on the brane assures that Lorentz invariance is maintained on the brane.
Therefore, momentum components are conserved on the brane, and we cannot generate
a nonzero $\dot{u}$ on the brane except as an initial condition.
Thus, we must have a nonzero $\dot{u}_0$ in order for the sterile neutrino to leave the brane and enjoy the bulk.
The uncertainty principle applied to the $u$ dimension
allows for such a nonzero velocity. In fact, it necessitates some uncertainty in $\dot{u}_0$.
With a brane thickness of $\Delta u$, one expects an initial $p_u=m\gamma\dot{u}_0 \sim 1/\Delta u$.
The path integral approach to quantum mechanics, a topic to which we will return later in this paper,
gives a similar result.  In the path integral approach, all $\dot{u}_0$ are allowed, with each value weighted by
the free-particle action $(S= m \int d\tau)/\hbar$.  The dominant paths are the classical extremae of the action,
and their width in action, of order $\hbar$, incorporates the quantum mechanical uncertainty.
Momentum conservation in the $u$-direction is a non-issue, as translational invariance
in the $u$-direction is broken by the brane itself;
alternatively, one may simply view the recoil of the brane as
compensating any momentum change in $p_u$.
\par
Since the set of geodesic equations are homogeneous in $\lambda$,
we may therefore replace $\lambda$ with $t$, and we do so.
Thus we arrive at two coupled geodesic equations,
    \be
        \frac{d}{dt} \left( e^{-2k |u|} \frac{dx}{dt} \right) &=& 0 \label{geo4}\\
        \frac{d^2u}{dt^2} \pm ke^{-2k |u|} \left(\frac{dx}{dt}\right)^2&=&0 \label{geo5}.
    \ee
For  Eq.~(\ref{geo4}) to be satisfied, the expression in parenthesis
must be a constant, independent of  the spacetime variables.
By taking $u$ to zero, it is clear that the constant is $\dxzero$, the initial velocity component
of the sterile neutrino along the brane.
So we have
    \be\label{geo6}
        \frac{dx}{dt}= \dot{x}_0\, e^{2k|u|}\,.
    \ee
Using this result, the geodesic Eq.~(\ref{geo5}) can be rewritten as
    \be
        \frac{d^2u}{dt^2} \pm k\,\dot{x}^2_0\, e^{2k|u|}=0 \label{geo7}.
    \ee
This has a standard form for a second order nonlinear differential equation,
which can be once-integrated  and recast as
    \be\label{geo8}
        \dot{u}^2-\dot{u}^2_0 = \mp 2 k\dot{x}^2_0 \int_0^{u(t)} du~e^{2k|u|}\,.
    \ee
\par
The trajectory of the sterile neutrinos is contained in
Eqs. (\ref{geo6}) and (\ref{geo8}).
From these equations, we may form the ratio $\dot{x}/\dot{u}=dx/du$
and solve for $x(u)$ or $u(x)$.
We obtain
    \be
        u(x)=\pm \frac{1}{2k} \ln\left[1+k^2x(l-x)\right] \label{u}\,,
    \ee
where $l=2 |\duzero| /k\dxzero$ is the distance along the brane at which the sterile neutrino
trajectory returns to the brane.
For small $kl\ll1$, this geodesic solution is the particularly simple parabola, $u(x)=\pm\frac{k}{2}\,x(l-x)$.
\par
Notice that the dimensionful warp parameter $k$ sets the scale for all length measurements.
Accordingly, Eq.~(\ref{u}) is especially simple when lengths are measured in units of $k^{-1}$.
\par
\begin{figure}
\centering
\raisebox{4.5cm}{${ku}$}
\includegraphics[scale=0.3]{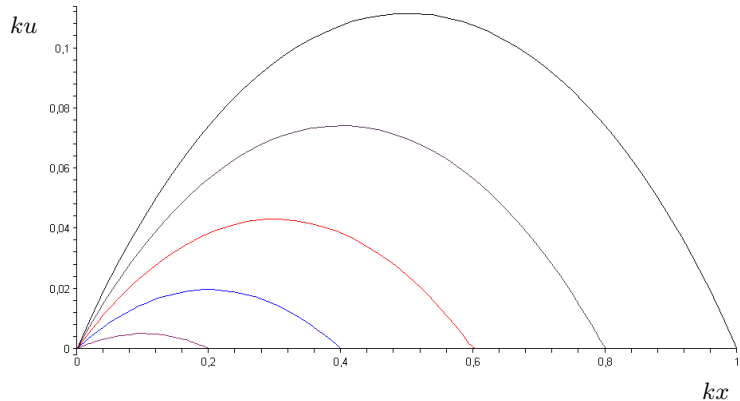}
\\{\hspace{9cm}${kx}$}
\caption{Sterile neutrino travel path in the extra dimension for
a travel ``distance'' $kl=0.2, 0.4, 0.6,0.8, 1$ as measured on the brane.} \label{fig:one}
\end{figure}

The geodesic is symmetric about the midpoint at $x=l/2$,
which implies that the angle of intersection of the sterile neutrino trajectory
coming back to the brane
is the same as the initial exit angle, given by $\arctan (\duzero/\dxzero)$.
From this we infer two features of the geodesic.
The first is that the exit angle is also equal to $\arctan (kl/2)$.
This presents a physical picture for the strength of the warp parameter $k$.
The second feature follows from the symmetry of the geodesic and the
``left-right $u$-symmetry" of the warp factor.
It is that $|u(x)|$ as given in Eq.~(\ref{u}) repeats itself
indefinitely, alternately on opposite sides of the brane.
Thus, the geodesic is periodic, with $l$ specifying the length of a half-cycle.
A half-cycle of the geodesic $ku (kx)$ is pictured in Fig.~\ref{fig:one} for some choices of $kl$.
The depth of penetration into the bulk is
    \beq\label{umax}
        \umax=|u(x=l/2)|=\frac{1}{2k}\ln \left[1+\left(\frac{kl}{2}\right)^2\right]\,.
    \eeq
This result provides another picture for the strength of the warp parameter.
At small $kl$ we have $ku_{\rm max} \approx (kl)^2/8$,
and at large $kl$ we have $ku_{\rm max}\approx \ln (kl/2)$.
\par
It will be useful to have expressions for the speeds $\duzero$ and $\dxzero$ in terms of
the geometric quantities $k$ and $l$.
From $l=2\duzero/k\dxzero$ above, and the constant of motion $\beta^2=\duzero^2+\dxzero^2$,
one easily finds
    \beq\label{dot01}
        \dot{x}_0=\frac{\beta}{\sqrt{1+\left(\frac{kl}{2}\right)^2}}\,,
        \qquad {\rm and} \qquad
        | \dot{u}_0 |=\frac{\beta\,(\frac{kl}{2})}{\sqrt{1+\left(\frac{kl}{2}\right)^2}}\,.
    \eeq
It is important to keep in mind that the distance $l$ as measured on the brane is a function of the initial values of the velocity components along the extra dimension and along the brane. Further in the paper we will mostly refer to dependencies on $l$ with the implicit dependence of $l$ on the initial conditions.
\par
\begin{figure}
\centering
\raisebox{5cm}{${\epsilon}$}
\includegraphics[scale=0.3]{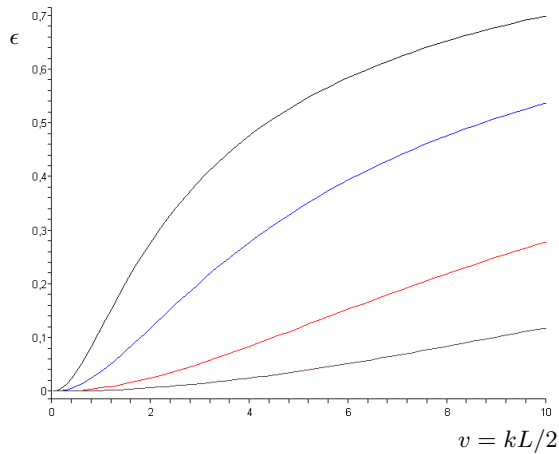}
\\{\hspace{6cm}${v=kL/2}$}
\caption{The shortcut parameter as a function of scaled baseline length $v=kL/2$.
Curves are parametrized by geodesic mode number $n=1, 2, 5, 10$ (from top to bottom).} \label{fig:two}
\end{figure}

The travel time $t^{\text{bulk}}$ of the sterile state in the bulk is
determined by substituting the expressions for $\dot{x}$ and $u(x)$ given in Eqs.~(\ref{geo6}) and (\ref{u}) into
    \be
        t^{\text{bulk}}(l)=\int_0^l \frac{dx}{\dot{x}(u(x))} =
        \int_{0}^{l} \frac{dx}{\dot{x}_0 [1+k^2x\,(l-x)]}
        =\frac{2}{\dot{x}_0 k\sqrt{1+\left(\frac{kl}{2}\right)^2}} ~
        \text{arcsinh}~\frac{kl}{2}
        = \left(\frac{l}{\beta}\right) \left(\frac{2}{kl} ~
        \text{arcsinh}~\frac{kl}{2}\right)\,.\label{tbulk}
    \ee
The travel time for the Standard Model
neutrinos confined to the brane is $t^{\text{brane}}(l)=l/\beta$ .
Thus, the sterile neutrino appears to a brane-bound observer to travel superluminally,
exceeding light-speed by the factor ($\frac{2}{kl}\,\text{arcsinh} \frac{kl}{2}$).
One can now write down the ``shortcut parameter" $\epsilon(l)$
in terms of the brane distance $l$ as
    \be
        \epsilon(l) \equiv  \frac{t^{\text{brane}}-t^{\text{bulk}}}{t^{\text{brane}}}=1-\left(\frac{\beta}{l}\right)\,t^{\text{bulk}}
           = 1-\left( \frac{2}{kl} \right)~\text{arcsinh}~ \left(\frac{kl}{2}\right)\,.\label{epsilon}
    \ee
An important feature has emerged.
The shortcut parameter depends on the length $l$ of the geodesic as seen from the brane.
According to Eq.~(\ref{umax}), the geodesic for large $l$ dives deeper into the bulk,
where the warp factor exponentially decreases the travel time.
The dependence of $\eps$ on $kl$ is shown in Fig.~\ref{fig:two}.
Focus here on the curve labeled $n=1$; the other curves will be explained shortly.
\par
Phenomenologically, we are interested in whether the sterile neutrino returns to the brane at a
fixed baseline $L$, where an experiment might be located.
If the sterile neutrino returns to the brane at $L$, then its wave packet will interfere with the
active neutrino packet to produce flavor oscillations.
If the sterile neutrino does not return to the brane at $L$, then there is no flavor interference.
The baseline $L$ and the return length $l$ are related in a simple way.
Given the periodicity of the geodesic, a sterile neutrino will interfere on the brane at $L$ if
$L$ is a multiple of the half-cycle length $l$, i.e., if
    \be�
        l=\frac{L}{n},\label{halfwavelength}
    \ee
where $n$ is any positive integer.
The travel time and therefore the shortcut parameter for a sterile neutrino which enters the detector
upon intersecting  the brane an $n^{th}$~time is simply obtained by
replacing $l$ with $L/n$ in Eqs.~(\ref{tbulk}) and (\ref{epsilon}), respectively.
The results are
    \be
       t^{\text{bulk}}_n (L) &=& \left( \frac{L}{\beta}\right)\,\left[ \left(\frac{n}{v}\right)\, \text{arcsinh} \left(\frac{v}{n}\right) \right]
       =  \left( \frac{L}{\beta}\right)\,(1-\eps_n)\,, \label{tbulkn} \\
       {\rm and\ } \qquad
       \epsilon_{n}(L) &=& 1-\left(\frac{n}{v}\right)\,
       \text{arcsinh}\left(\frac{v}{n}\right)\,.   \label{epsilonn}
    \ee
Here we have defined the dimensionless ``scaling variable"
    \be
        v\equiv \frac{kL}{2}\,\label{v}.
    \ee
This variable will appear throughout subsequent equations.
Its dimensionless feature proves useful when a perturbation series in $v$ or $v/n$ is desired.
Fig.~\ref{fig:two} shows the dependence of $\epsilon_{n}(L)$ on $v=kL/2$ and $n$.
It is seen that $\eps$ grows monotonically with $v$, and for fixed $v$, shrinks monotonically with
geodesic mode number $n$.
Since we have seen in Eq.~(\ref{Eres})
that the resonance energy $\Eres$ varies as $1/\sqrt{\eps}$, the energy
$\Eres$ will shrink monotonically with increasing $L$, and grow monotonically with increasing $n$.
\par
Anticipating the next sections of this paper, we expand Eq.~(\ref{epsilonn}) in $v/n$.
The result is
    \be
        \epsilon_{n}(L) = \frac{1}{6}\,\left(\frac{v}{n}\right)^2\,
        \left[1-\frac{9}{20}\left(\frac{v}{n}\right)^2+\cdots\right]\,.\label{smallepsilon}
    \ee
The leading behavior tells us that small $\eps$ and small $v^2$ are synonymous,
and that $\eps_n\approx\frac{1}{24}\,(kL/n)^2\propto L^2$, for small $v/n$.
Therefore, for small $v/n$ one has
$\Eres (n)\approx 2n\sqrt{3\Delta m^2\,\cos 2\theta}/kL$ as the resonant energy of the $n^{th}$ mode.
\par
Also useful will be the generalization of Eq.~(\ref{dot01}), obtained by setting $l=L/n$.
The generalization is
    \beq\label{dot02}
        \dot{x}_0=\frac{\beta}{\sqrt{1+\left(\frac{v}{n}\right)^2}}\,,
        \qquad {\rm and} \qquad
        |\dot{u}_0|=\frac{\beta\,\left(\frac{v}{n}\right)}{\sqrt{1+\left(\frac{v}{n}\right)^2}}
        =\frac{\beta\,v}{\sqrt{n^2 +v^2}}\,.
    \eeq
Eq.~(\ref{dot02}) reveals that the initial angle between the sterile trajectory and the brane,
$\alpha_0$, is given by
$\tan\alpha_0= |\duzero|/\dxzero = v/n$.

\section{Summing the geodesic paths}\label{sec:summing}

Each geodesic, labeled by integer $n$'s, presents a possible path for the sterile neutrino.
According to quantum mechanics, we must sum over these possibilities (amplitudes $A(n)$),
and square the total amplitude ($A=\sum_n A(n)$).
In this way, our calculation becomes a semi-classical approach to path-integral quantum mechanics.
Each geodesic is a classical path, and so is also an extremum of the path integral.
\par
What is absent at this point is the weight to be associated with each classical path.
There are actually three weights to be considered.

\subsection{Path-integral weight} \label{subsec:qmweight}

The first is the usual quantum mechanical weight
$e^{iS_n}$, where
    \be
        S_n=m\int d\tau = \frac{m^2}{E}\int dt = \left(\frac{m^2}{E}\right)\,t^{\text{bulk}}_n
        = \left(\frac{m^2 L}{\beta E}\right)\,(1-\eps_n)\label{sn}
    \ee
is the $n^{th}$ mode action for a free-particle.
The variables $t^{\text{bulk}}_n$ and $\eps_n$ are given in Eqs.~(\ref{tbulkn}) and (\ref{epsilonn}) above.
The small $v$ expansion of $S_n$ is
    \be\label{actionsmallv}
        S_n=\left( \frac{m^2 L}{\beta E} \right) \, \left[ 1-\frac{1}{6} \left(\frac{v}{n}\right)^2 +\cdots \right] \,.
    \ee
To leading order, $S_n$ is independent of $n$.
Therefore, for small $v/n$, the unimodular weight $e^{iS_n}$ can be omitted from the sum.
This means that for small $v$, all amplitudes $A(n)$ are relatively real and so interfere constructively.
On the other hand, for intermediate and large values of $v/n$,
the weight is $n$-dependent and must be retained.
For intermediate and large $v/n$,
interferences among modes depend on the phase $S_n$.
However, from Eq.~(\ref{Eresn})
we infer that $\Eres^2 (n) = \Delta m^2\,\cos 2\theta /2\,\eps_n$,
so we may infer that for a resonant energy in the interesting range
$\Eres (n) \gg \sqrt{\Delta m^2}$, we are in the small $\eps_n$, equivalently, small $v/n$ domain,
and the weight $e^{i S_n}$ can be omitted.

\subsection{Distribution of initial \boldmath{$\duzero$}}\label{subsec:duzero}

The second weight concerns the {\it a priori} amplitude for nonzero values of $\duzero$.
According to Eq.~(\ref{dot02}), there is a one-to-one correspondence between
$v/n=kL/2n$ and $\duzero$.
Thus, each mode intersecting the brane results from a unique initial value of $\duzero$.
Here we make our most model-dependent assumption, namely that the initial amplitudes
in momentum $p_u=m\gamma\duzero$ transverse to the brane
assume a normal distribution about $p_u=0$.
The arguments favoring this choice are standard:
Ground state wave functions often have a Gau{\ss}ian distribution about the mean;
random fluctuations are distributed in Gau{\ss}ian form;
while initial momenta may prefer the brane direction, the uncertainty principle
requires a nonzero $p_u$ component as well.
Thus, we assume for the $\duzero$-distribution the following normalized Gau{\ss}ian
    \be\label{dNG0}
        dN_G (p_u)=\frac{1}{\sqrt{2\pi}\sigma}
        ~e^{-\left\{ \frac{p_u^2}{2\sigma^2} \right\} }~dp_u\,,
    \ee
which satisfies the usual normalization condition
    \be
        \int_{-\infty}^{+\infty}\frac{1}{\sqrt{2\pi}\sigma}
        ~e^{ -\left\{ \frac{p_u^2}{2\sigma^2} \right\} }~dp_u = 1 \,.\label{normalization}
    \ee
The new parameter $\sigma$ is the rms value of $p_u$.
Heisenberg's uncertainty principle relates $\sigma$ to the uncertainty in the initial $u$-position $\Delta u$ via
    \be
        \sigma \geq \frac{1}{2 \Delta u}.
    \ee
We will take $\sigma\sim 1/\Delta u$, and interpret the uncertainty $\Delta u$ as some measure of the thickness of our brane.
Another estimate would be to set $\sigma$ equal to the only extra-dimensional scale in the problem, the warp parameter $k$.
If the extra dimensions are compactified at a length scale $\mathcal M_d^{-1}$, then certainly
$\Delta u <  \mathcal M_d^{-1}$, or equivalently, $\sigma > \mathcal M_d$.
Reference~ \cite{Hoyle:2004cw} provides a lower bound on the scale of the extra dimension $\mathcal M_d \simeq  10^{-3}~ \text{eV}$,
or $\mathcal M_d^{-1} \simeq  0.2$~mm.
While $\sigma\sim 1/\Delta u$ is a parameter which in principle can be fit to data, e.g. to the height of the resonance oscillation peak, as we show later.
We will use a fiducial scale of one micron for $\Delta u$ in our numerical work.
This translates into a value of $0.2$~eV for the rms momentum $\sigma$,
which is coincidentally (or not?) of order of the scale of neutrino masses.
\par
Employing $m\gamma =E$ and Eq.~(\ref{dot02}),
the momentum component along $u$ can be written  as
    \be
        p_u = \gamma m\duzero=\frac{E \beta v}{\sqrt{n^2+v^2}}\,.\label{relmom1}
    \ee
Substituting for $p_u$ with Eq.~(\ref{relmom1}),
our differential $\duzero$-distribution in Eq.~(\ref{dNG0}) can be rewritten as
    \be
        dN_G (n)&=&\frac{1}{\sqrt{2\pi}\,\sigma}
        ~e^{-\left\{ \frac{(E\beta v)^2}{2\sigma^2(n^2+v^2)} \right\} }~\frac{dp}{dn}~dn \nonumber \\
        &=&\frac{1}{\sqrt{2\pi}\,\sigma}~e^{ - \left\{ \frac{(E\beta v)^2}{2\sigma^2(n^2+v^2)} \right\} }
        ~\frac{E\beta v n}{(n^2+v^2)^{3/2}}~(-dn). \label{dNG1}
    \ee
Integration of $(-dn)$ is to be interpreted as
    \beq\label{Deltan}
        2\int_\infty^0 (-dn) = 2\int_0^\infty dn = 2\sum_{n=1}^\infty\, \Delta n\,,
    \eeq
where the factor of $2$ accounts for the negative integers which label the geodesics
whose first half-cycle is in the negative $u$ half-plane,
and $\Delta n$ is a measure to be associated with each geodesic mode.
Thus, we arrive at
    \be
        \int dN_G (n)&=&\sqrt{\frac{2}{\pi}}\, \frac{E\,\beta}{\sigma} \sum_{n=1}^\infty \Delta n ~\frac{v\,n\ \ \ }{(n^2+v^2)^{3/2}}
        ~e^{ -\left\{ \frac{(E\beta v)^2} {2\sigma^2(n^2+v^2)} \right\} }\,. \label{dNG}
    \ee
If $\Delta n$ were unity for all $n$, then the sum amounts to approximating the Gau{\ss}ian integral
by a histogram, with each integer specifying a bin of unit width.  This would yield
$\sum_n \frac{dN}{dn} \approx 1$, which means that every neutrino which leaves the
brane appears at $x=L$.  Since in general some sterile neutrinos leaving the brane
will not arrive at $x=L$, this choice of measure for $\Delta n$ cannot be so simple.
The measure for $\Delta n$, our third weight, will be discussed shortly.
\par
Guesses for the initial distribution of $\duzero$'s other than the Gau{\ss}ian form presented above are possible.
A simple example would be to distribute $\duzero$ uniformly over the interval $[-\bu,\ \bu]$.
With this guess, one has the flat distribution
    \beq dN_F(n) =\left( \frac{dN_F}{d\duzero} \right) \left( \frac{d\duzero}{dn} \right) \,dn
        =\left(\frac{1}{2\bu}\right) \left(\frac{\beta\,v\,n}{(n^2 +v^2)^{3/2}}\right) (-dn)\,.
    \eeq
Employing the left-right bulk symmetry of our metric and moving to discrete notation,
this may be written as
    \beq \label{dNF}
        \int dN_F (n)= \frac{\beta}{\bu}
           \sum_{n=1}^\infty \Delta n \; \frac{v\,n\ \ \ }{(n^2 +v^2)^{3/2}}\,.
    \eeq
\par
Similarities and differences between Eqs.~(\ref{dNG}) and (\ref{dNF}) are evident.
The advantages of the $\duzero$-distribution in Eq. (\ref{dNG}) are that the Gau{\ss}ian form is motivated by
physical arguments given earlier, and that the distribution is sensitive to the brane thickness
$\Delta u \sim 1/\sigma$.  One expects the brane thickness to enter into the physics, and it does so here.
A disadvantage for Eq. (\ref{dNG}) is that there is one more parameter assignment to be made, the value for $\sigma$.
However, since $\sigma$ entered originally as part of the normalization of the continuous $\duzero$-distribution, and therefore vanished upon integration,
we do not expect much residual sensitivity to $\sigma$ when we move to the discrete weighted sum.
The advantage of the $\duzero$-distribution in Eq. (\ref{dNF}) is its sheer simplicity.  This distribution may be
motivated by appealing to path-integral quantum mechanics, where {\sl a priori} all initial velocities are
allowed with equal weight.
In fact, we may write both distributions $dN_G$ and $dN_F$ in a similar way:
    \beq\label{distributions}
        \int dN_D (n) = \sum_{n=1}^\infty \Delta n\;\left( \frac{n}{n^2+v^2}\right)
        \left \{
        \begin{array}{ll}
        \sqrt{\frac{2}{\pi}}\,\left | \frac{d}{dz} \left(e^{-z^2/2}\right) \right |_{
        z=\left( \frac{\beta E}{\sigma}\frac{v}{\sqrt{n^2+v^2}} \right) }\,,
        & D=G{\rm au{\ss}ian} \\
        & \\
        \frac{\beta}{\bu}  \frac{v}{\sqrt{n^2+v^2}} \,, & D=F{\rm lat} \, . \\
        \end{array}
        \right.
    \eeq
Written in this form, the similarities and differences are more apparent.
Both bracketed expressions are bounded from above by ${\cal O} (1)$.
In fact, the slope of the Gau{\ss}ian is bounded by $1/\sqrt{e}\sim 0.7$,
and both are small for small $v$.
The differences appear at very small $v$ and at large $v$.
At very small $v$, the bracketed expression is simply $v/n$ for $dN_F$, while
for $dN_G$ it is $\sqrt{2/\pi}\,\beta E v/\sigma n$,
and so is dependent on the brane width $\Delta u\sim1/\sigma$ for $dN_G$.
At resonance, the bracketed expression for $dN_G$ is $\sqrt{6\Delta m^2 \cos 2\theta/\pi}\,(\beta/\sigma)\sim m/\sigma$.
Also, $dN_G$ is small for large $v$ (the tail of the Gau{\ss}ian),
whereas $dN_F$ is ${\cal O}(1)$ at large $v$.
As mentioned above, for a resonance to occur in an interesting energy range requires that $v$ be small,
so small $v$ becomes our focus.
\par
We will analyze the implications of both $\duzero$-distributions.
For the Flat $\duzero$-distribution, we will set $\bu=\beta$ in our numerical work.

\subsection{The weight of \boldmath{$\Delta n$}} \label{subsec:Deltan}

We come to the third and final weight, $\Delta n$,
characterizing the thickness of the classical path in $n$-space.
According to path-integral quantum mechanics,
the thickness of the path is determined by the variation of $e^{iS/\hbar}$ about
the classical extremum $e^{iS_{cl}/\hbar}$.  For $| S-S_{cl} |\agt \hbar$, variations are rapid
and the integration averages to zero. Thus, $\Delta S \sim \hbar$ characterizes the thickness
of the path. So we have
    \beq\label{Deltan2}
        \Delta n = \frac{dn}{dS}\,\Delta S \sim \frac{1}{dS/dn}\,,
    \eeq
where $S(n)$ is viewed as a continuous function of $n$.
Non-integer values of $n$ describe geodesics which interpolate between the integer-labeled modes,
i.e. they describe classical paths which do not intersect $x$ at $L$.
But if the dependence of $S$ on $n$ is mild, i.e., if $dS/dn$ is small, then the quantum mechanical
uncertainty allows the interpolated paths to intersect $x$ at $L$, too.
$S_{cl}$ is the extremum of the multi-functional $S(t;x,u)$ that characterizes a classical path.
But $S_{cl}$ is not an extremum of $S$ in the variable $n$.  We assume the dependence of
$S$ on $n$ is sufficiently mild that $dS/dn\sim dS_{cl}/dn$.
From Eqs.~(\ref{sn}) and (\ref{epsilonn}) we easily obtain
    \be
        \frac{d S_{cl}(n)}{dn}=\frac{2m^2}{E \beta k} \left(\text{arcsinh} \frac{v}{n}-\frac{v}{\sqrt{v^2+n^2}}\right).
    \ee
Of course, $\Delta n$ cannot exceed unity, the distance between successive integers.
So we arrive at our definition of the measure
    \be
        \Delta n=\text{min}\left\{
        1,\frac{E \beta\,v}{L m^2}
        \left(\text{arcsinh} \frac{v}{n}-\frac{v}{\sqrt{v^2+n^2}}\right)^{-1}
        \right\}\,.   \label{Deltan3}
    \ee
The small $v$ expansion of Eq. (\ref{Deltan3}) is
    \be
        \Delta n=\text{min}\left\{
        1,\frac{E \beta}{L m^2} \left(\frac{3\,n^3}{v^2}\right)
        \left[ 1+\frac{9}{10}\left(\frac{v}{n}\right)^2+\cdots \right]
        \right\}\,, \quad{\rm for\ } v^2\ll 1\,.   \label{Deltansmallv}
    \ee
\par
It is useful at this point to introduce the usual vacuum oscillation length for
neutrinos on the 4D brane $\lamosc=4\pi\,E/\Delta m^2=2.48\,{\rm km} (E/{\rm GeV}) (\Delta m^2/{\rm eV}^2)^{-1}$.
Then the prefactor $E \beta /L m^2$ may be written as $\beta/ 4\pi\,\Nlam$,
assuming $m^2 \gg m_a^2$, and
where we have introduced the dimensionless
    \beq
        \Nlam = \frac{L}{\lamosc} =\frac{L\Delta m^2}{4\pi E},
    \eeq
which is equal to
the number of vacuum oscillations in the baseline interval $0$ to $L$.
$\Nlam$ is a useful variable in that it is effectively bounded by 100, at most,
if oscillations are to occur.
This bound arises from ``experimental decoherence'' as follows~\footnote
{
There is also the possibility of wave packet decoherence, which arises when the two mass states in the wave function
separate by more than the length of the wave packet. The condition for wave function decoherence
is $\delta x=\eps L >\delta\Psi$, i.e., $\eps> \delta\Psi/L$,
where $\delta\Psi$ is the length of the wave packet.
The length of the wave packet is determined by the time-scale governing the production of the neutrino.
For example, neutrinos resulting from $\pi^{\pm}$-decay would have
$\delta\Psi\sim c\gamma\tau_{\pi^\pm}\sim (E_\pi/m_\pi)\times 0.3$~km.
Only for astrophysical distances does wave packet decoherence become significant,
and we consider it no further in this paper.
}:
The 4D~vacuum oscillation phase is $\phi=L\,\Delta m^2/2\,E$.
Therefore, experimental uncertainties in $L$ and $E$ introduce an uncertainty in the phase of
    \beq
        \delta\phi= \phi\,\sqrt{ \left(\frac{\delta L}{L}\right)^2 + \left( \frac{\delta E}{E} \right)^2 }\,.
    \eeq
But $\phi=2\pi\,\Nlam$, so even if the relative experimental uncertainties in $L$ and $E$
were as small as a few percent, the phase uncertainty becomes total, i.e. $\delta\phi=2\pi$, before $\Nlam$ reaches 100.
So by writing the explicit $L\,m^2/E$~dependence in terms of the dimensionless
$\Nlam$, we are assured of a pure number whose value cannot exceed 100 if coherence is
to be maintained.
Thus, the option $\frac{E \beta}{L m^2} \left(\frac{3\,n^3}{v^2}\right)$  in Eq.~(\ref{Deltansmallv})
is greater than $\sim 3\,n^3/ (400\pi\,v^2)$ if coherence is to be maintained.
This exceeds unity, the alternate option,
if $v^2 \alt n^3/400$.
So for $v \alt n^{3/2} /20$, if coherence is to be maintained, then we have $\Delta n (n) =1$.
If $v \alt 1/20$, then $\Delta n (n) =1$ for all $n$.
The physics of this result is that for  $v < n^{3/2} /20$,
the action is sufficiently flat in $n$
(Recall that $S_n$ as given in Eq.~(\ref{actionsmallv}) is nearly independent of $n$ for small $v/n$.)
that all neutrinos leaving the brane on trajectories labeled by
the interval $[n,\,n+1]$ return to the brane at $x=L$~\footnote
{
We could make a more careful evaluation of $\Delta n$ by enlarging the return point $(x,u)=(L,0)$
to an area defined by the brane thickness $\Delta u$ and the detector size $\Delta x$,
but such additional detail seems unwarranted for the present project.
}.
\par
Eqs.~(\ref{umax}) and (\ref{v}) give the maximum classical penetration depth into the bulk as
    \beq\label{umax2}
        \umax=\frac{L}{4v} \ln\left[ 1+\left(\frac{v}{n}\right)^2\right]\,,
    \eeq
which has a small $v/n$ expansion
    \beq\label{umaxsmall}
        \umax=\frac{L\,v}{4n^2}\,\left[ 1-\frac{1}{2}\left(\frac{v}{n}\right)^2+\cdots \right]\,.
    \eeq
Thus we see again that for small $v$, the geodesics never wander farther than $vL/4$ from the brane.
In this sense it is not surprising that all neutrinos leaving the brane at small angles
$\arctan v \sim v$ have an amplitude to be found on the brane at $x=L$.
\par
If $\Nlam$ does exceed ${\cal O}(100)$, coherence is lost, oscillations are unobservable,
but oscillation-averaged mixing persists as the vestigial signature of oscillations.
In fact, it is $\NlamFiveD$ which is relevant for determining coherence or decoherence in the
present model.  Setting $\lamoscFiveD=2\pi/\delta\tilde H_n$ leads to the relation
\be
\NlamFiveD=\Nlam\,\sqrt{\sin^2 2\theta + \cos^2 2\theta \left[1-\frac{E^2}{\Eres^2}\right]^2}.
\ee
Except near a resonance, $\Nlam$ is a good guide for $\NlamFiveD$.
At a resonance, the value of $\Nlam$ overestimates the number of oscillations by the factor of $1/\sin 2\theta$.
We return to the possibility of decoherent mixing in the next subsection.

\subsection{Putting all the pieces together}\label{subsec:together}

The total amplitude for the active-sterile neutrino oscillations is the sum of the amplitudes $A(n)$ for each mode $n$.
Each $A(n)$ includes the three weights $e^{iS_{cl}(n)}$, $dN_D$, and $\Delta n$.
We have
    \be
        A_{as} = \sum_{n=1}^\infty A(n)
        = \sum_{n=1}^{\infty} \Delta n\;e^{iS_{cl}(n)}\,
        \frac{v n}{(n^2+v^2)^{3/2}}~
        \left[  \sqrt{\frac{2}{\pi}}\,\frac{\beta E}{\sigma}~ e^{-\frac{(\beta E v)^2}{2\sigma^2(n^2+v^2)}}  \right]
        \sin 2\tilde\theta_n \ \sin  \frac{L\delta\tilde{H}_n}{2}\,,
        \label{amplitude}
    \ee
with $\sin 2\tilde\theta_n$ and $\delta\tilde H_n$ obtained from $\sin 2\tilde\theta$ and $\delta\tilde H$
in Eqs.~(\ref{tildesin}) and (\ref{tildeH}) by replacing $\Eres$ with the resonant energy of the mode		
    \beq\label{Eresn}
        \Eres (n) = \sqrt{\frac{\Delta m^2 \cos2\theta}{2\eps_n}}
        = \sqrt{\frac{\Delta m^2 \cos2\theta}{2\left[1-\left(\frac{n}{v}\right)
        \text{arcsinh}\left(\frac{v}{n}\right)\right]}}  \,.
    \eeq
The bracketed expression in Eq.~(\ref{amplitude}) pertains to the $dN_G$~weighting of the $\duzero$-distribution.
For the $dN_F$~weighting, this bracketed expression is replaced by $\beta/\bu$.
The probability of oscillation is the square of the amplitude of oscillation,
    \beq\label{probas}
        P_{as}=|A_{as}|^2 = \left|\sum_{n=1}^\infty A(n) \right|^2.
    \eeq
\par
Recall that if $\Nlam$ exceeds ${\cal O}(100)$, coherence is lost and so oscillations are unobservable.
Phase-averaging then sets $\langle \sin\frac{L\delta\tilde H_n}{2}\rangle$ to zero
and $\langle \sin^2\frac{L\delta\tilde H_n}{2}\rangle$ to $\frac{1}{2}$.
Implementing this averaging in Eq.~(\ref{probas}) then leads to the classical expression
    \beq\label{mixas}
        \langle P_{as}\rangle_{\rm phase-ave}=  \sum_{n=1}^\infty \langle |A(n)|^2\rangle
        = \frac{v^2}{2} \sum_{n=1}^\infty  (\Delta n)^2 \, \frac{n^2}{(n^2+v^2)^3} \, \sin^2 2\tilde\theta_n
           \left[ \frac{2}{\pi} \left(\frac{\beta E}{\sigma}\right)^2 \,e^{-\frac{(\beta E v)^2}{\sigma^2(n^2+v^2)}}\right] \,.
    \eeq
As with the quantum mechanical amplitudes in Eqs.~(\ref{amplitude}) and (\ref{probas}),
the bracketed expression pertains to the $dN_G$~weighting,
while the analogous result for the model with $dN_F$ weighting is obtained by setting the bracketed expression to $(\beta/\bu)^2$.  As we showed earlier, near a resonance the coherence length is increased
by a factor of $1/\sin 2\theta$.  Thus, the phase-averaged Eq.~(\ref{mixas}) may not apply
if $LE$ lies near a resonance.

\section{Some Limiting Cases}
\label{sec:limitingcases}
In this section we examine vacuum limits in 4D, and in our higher-dimensional 5D~model.
Vacuum limits are the expressions  for oscillation probabilities that are valid at energies well
below any resonance(s).
The vacuum limits are different for the 5D~model, and the standard 4D~model.

\subsection{The 4D~Vacuum Limit}
\label{subsec:4Dlimit}
The standard 4D~vacuum expression for the two-state oscillation probability is the familiar one,
\beq\label{4Dvacuumprob}
P_{as}({\rm 4D~vacuum})=\left| \sum_j U_{sj} e^{-i\frac{Lm^2_j}{2E}} U^*_{aj}\right|^{\ 2}
	= \sin^2 2\theta\,\sin^2 \frac{L\Delta m^2}{4E}\,.
\eeq

\subsection{The 5D~Vacuum Limit}
\label{subsec:5Dlimit}
At $LE\ll \ELres$ (the notion of a resonant product $\ELres$ will be discussed in section \ref{sec:nearzone}),  $\sin L\delta\tilde H/2$ and $\sin^2\tilde\theta$ given in Eqns.~(\ref{tildeH}) and (\ref{tildesin}), respectively, assume standard 4D vacuum values.  Thus, from Eqs.~(\ref{probas}) and (\ref{amplitude}) we have
\beq\label{5Dvacuumprob}
P_{as}({\rm 5D~vacuum}) = P_{as}({\rm 4D~vacuum}) \times
	\left| \sum_{n=1}^{\infty} \Delta n\;e^{iS_{cl}(n)}\,
        \frac{v n}{(n^2+v^2)^{3/2}}~
        \left[  \sqrt{\frac{2}{\pi}}\,\frac{\beta E}{\sigma}~ e^{-\frac{(\beta E v)^2}{2\sigma^2(n^2+v^2)}}  \right]\,\right|^{\ 2}
\eeq
for the Gau{\ss}ian $\duzero$-distribution,
and the bracketed expression replaced by $\beta/\bu$ for the Flat $\duzero$-distribution.

Clearly, the 5D~model is richer than the 4D~model, and more complicated.	
Nevertheless, for consistency of our 5D~model, there must exist a limit in which the standard four-dimensional vacuum solution is obtained.  Next we present this limit.

\subsection{The 4D Limit of the 5D Model}
\label{subsec:5Dto4D}
Since in 4D all of spacetime comprises our brane, in the 4D~limit the sterile neutrino,
like the active neutrinos, is confined to the brane.
Thus, the sterile neutrino exit angle $\alpha_0=\arctan v/n$ must be set to zero.
This implies that the warp factor $k$ must be set to zero.
Equivalently, we may take $|\duzero|$ to zero.
According to Eq.~(\ref{dot02}), this again implies that $k$ must go to zero.
From Eq.~(\ref{epsilonn}) we see that $\eps_n$ equals zero for all $n$ in the $k=0$ limit.
The vanishing of $\eps_n$ in turn implies that $\Eres$ given by Eq.~(\ref{Eresn}) is infinite, and so
$\delta\tilde H$ and $\sin^2\tilde\theta$ given in Eqns.~(\ref{tildeH}) and (\ref{tildesin}), respectively,
assume standard 4D vacuum values.
At this point, the 4D~limit of the 5D~probability has the form given in Eq.~(\ref{5Dvacuumprob}).
\par
Further simplification occurs.
From Eq.~(\ref{Deltansmallv}) we see that $\Delta n$ is equal to unity in the $k=0$ limit.
Furthermore, the vanishing of $\eps_n$ and Eq.~(\ref{sn}) together reveal that
$e^{iS_{cl}(n)}$ is an irrelevant, $n$-independent, unimodular constant.
\par
We are left to consider the mode integration $\int dN_{D}$, which must go to unity
(one trajectory for the brane-bound sterile neutrino) in the limit $k\srarr 0$.
In the Gau{\ss}ian ($D=G$) case, we must take $\sigma\srarr 0$ to reduce the $p_u$-distribution
to a delta function centered at $p_u=0$, while in the Flat ($D=F$) case, we must take $\bu\srarr 0$
to reduce the $\duzero$-distribution to a delta function at $\duzero=0$.
The zero limits of $\sigma$ or $\bu$, and $k$, must be coordinated such that the mode integrals
in Eqs.~(\ref{dNG}) and (\ref{dNF}) go to unity in these limits.
Let us define the ratios $r_G\equiv k/\sigma$ and $r_F\equiv k/\bu$.
From Eqs.~(\ref{dNG}) and (\ref{dNF}), we therefore require that
\beq\label{modereduction1}
1=\sum_{n=1}^{\infty} \frac{1}{n^2}
 \left\{
	\begin{array}{lll}
	\sqrt{\frac{2}{\pi}} \left(\frac{\beta ELr_G}{2}\right) e^{-\frac{1}{2}\left(\frac{\beta ELr_G}{2n}\right)^2}\,, &
	\quad {\rm Gau{\ss}ian\ distribution}\,\\
 & \\
	\frac{\beta Lr_F}{2}\,, & \quad {\rm Flat\ distribution} \,. \\
	\end{array}
\right.
\eeq
Thus, $\sigma$ goes to zero as $k/r^*_G$, and $\bu$ goes to zero as $k/r^*_F$,
where $r^*$ is the solution of Eq.~(\ref{modereduction1}).
For the Gau{\ss}ian case, Eq.~(\ref{modereduction1}) is transcendental.
However, for the Flat case, the solution is simply $r^*_F=2/\zeta(2) \beta L$, where
$\zeta(2)\equiv \sum_{n=1} 1/n^2 = \pi^2/6$.
The recipe for reduction of our model to the usual 4D~vacuum solution is now complete.
One takes $k\srarr 0$ and $r\srarr r^*$.
\par
Notice that $n$ is an irrelevant parameter for motion on the brane.
Accordingly, there is another, simpler, approach connecting our model to the 4D~vacuum result.
One may omit the sum on $n$ and just set $n$ equal to one.
Then, the $r^*$'s are solutions to the simpler normalization equations
\beq\label{modereduction2}
1=
\left\{
	\begin{array}{lll}
	\sqrt{\frac{2}{\pi}} \left(\frac{\beta ELr_G}{2}\right) e^{-\frac{1}{2}\left(\frac{\beta ELr_G}{2}\right)^2}\,, &
	\quad {\rm Gau{\ss}ian\ distribution}\,  \\
 & \\
	\frac{\beta Lr_F}{2}\,, & \quad {\rm Flat\ distribution} \,. \\
	\end{array}
\right.
\eeq
In particular, $r^*_F (n=1) = 2/\beta L$ with this recipe.
\par
It should be noted that although our model in extra-dimensional warped space has a 4D~vacuum limit,
the extra-dimensional model is much richer in phenomenology.  It includes two new physically-motivated parameters, the warp parameter $k$, and the $\duzero$-distribution parameter $\sigma$
or $\bu$.  One consequence of the dramatic difference is that there is no reason to expect the 4D~vacuum parameters $\theta$ and $\Delta m^2$ to correlate in value with the same parameters in our model.

\section{The near zone}\label{sec:nearzone}
Of particular interest is the ``near zone''', which we define as
    \beq\label{nearzone}
        v \alt \frac{1}{2}\,, \qquad {\rm or\ equivalently,}\qquad kL \alt 1 \qquad [``{\rm Near\ Zone}"]\,.
    \eeq
The near zone is interesting for two related reasons.
The first is that it is here in the near zone where an experiment can hope to see resonant behavior.
From Eq.~(\ref{Eresn}) we have that
$\Eres^2 (n) /\Delta m^2 = \cos 2\theta/2\eps_n$,
so a resonant energy well above the sterile neutrino mass requires a small $\eps_n$,
as noted earlier in Eq.~(\ref{smalleps}) and~\cite{Pas:2005rb}.
From the form of $\eps (v/n)$ given in Eq.~(\ref{smallepsilon})
or from Fig.~\ref{fig:two} we have seen that small $\eps_n$ means small $v/n$,
exactly the condition employed to define the near zone.
\par
The second reason the near zone is interesting is that our formulas become especially simple in
the near zone.
The value of $\eps_n$ at small $v/n$, given in Eq.~(\ref{smallepsilon}) is $v^2/6n^2$,
which fixes the resonant energies, for $\Eres^2 (n) \gg \Delta m^2$, to be
    \beq\label{Eresnearzone}
        \Eres (n) = \frac{n}{kL}\sqrt{12\,\Delta m^2\,\cos 2\theta}\,.
    \eeq
Something very interesting has emerged.
The two parameters under experimental control are the energy $E$ and the baseline $L$.
We define the ``resonant product of $E$ and $L$" as
    \beq\label{ELres}
        \ELres \equiv \frac{1}{k}\,\sqrt{12 \Delta m^2 \cos 2\theta}\,.
    \eeq
Then we have that
    \beq\label{Esqratio}
        \frac{E^2}{\Eres^2 (n)}= \frac{(LE)^2}{n^2\,\ELres^2}
    \eeq
in the near zone.
Substituting this result into the generalizations of Eqs.~(\ref{tildesin}) and (\ref{tildeH}) to the geodesic modes, we obtain
    \be
        \delta\tilde H_n = \frac{\Delta m^2}{2E}\sqrt{\sin^22\theta + \cos^22\theta
        \left[1-\frac{(LE)^2}{n^2\,\ELres^2}\right]^2} \label{deltaHofLE}
    \ee
and
    \be
        \sin 2\tilde\theta_n = \frac{\sin 2\theta}{\sqrt{ \sin^2 2\theta + \cos^2 2\theta
        \left[1-\frac{(LE)^2}{n^2\,\ELres^2} \right]^2}} \,.  \label{sinthetaofLE}
    \ee
It is now clear that in the near zone, where $\Eres^2\gg\Delta m^2$ or equivalently
$kL\ll 1$, the location of the resonance depends not just on $E$ but also on $L$,
in the very simple combination $LE$.
\par
The amplitude for the $n^{th}$ mode contains the product of sines
\beq\label{sineprod}
\sin 2\tilde\theta_n\,\sin \frac{L\delta \tilde H_n}{2}
	\ \ \stackrel{LE=n\ELres}{\longrightarrow}
	\ \ \sin\left(\pi\Nlam\sin 2\theta\right)\,,
\eeq
where the rhs is the maximum value of the product, attained whenever a resonant value of $LE$ is reached,
in which case the square root in Eq.~(\ref{deltaHofLE}) or (\ref{sinthetaofLE}) is minimized.
Note that the resonant value is independent of $n$, and is bounded in magnitude by unity.
Thus, this product cannot contribute significantly to the convergence of the sum, which we will discuss shortly.
\par
Further simplification of formulas appear in the near zone.
The expansion of $\eps_n$, $S_n$ and $\Delta n$ in powers of small $v/n$ are already given
in Eqs.~(\ref{smallepsilon}), (\ref{actionsmallv}), and (\ref{Deltansmallv}), respectively.
Eqs.~(\ref{distributions}) and (\ref{Deltansmallv}) yield the small $v$ limits for the $\duzero$-distributions $\int dN_G$ and $\int dN_F$.
The ultimate result is the near zone limit for the amplitude of Eq.~(\ref{amplitude}),
which to leading nonzero order in $v^2$ is, for the case of the Gau{\ss}ian $\duzero$-distribution,
    \beq
        P_{as}  = \left|
        \sum_{n=1}^{\infty} \Delta n (n)\,
        \left( \frac{v}{n^2}\right)
        \left[  \sqrt{\frac{2}{\pi}}\,\frac{\beta E}{\sigma}~ e^{-\frac{1}{2}\left(\frac{\beta E v}{\sigma n}\right)^2}  \right]
        \sin 2\tilde\theta_n \ \sin \frac{L\delta\tilde H_n}{2} \right|^{\ 2}\,,
        \label{probnearzone}
    \eeq
where $\sin 2\tilde\theta_n$ and $\sin (L\delta\tilde{H}_n/2)$
are given by the their near zone expressions in Eqs.~(\ref{sinthetaofLE}) and (\ref{deltaHofLE}),
which show the explicit dependence on the resonant product $\ELres$ defined in Eq.~(\ref{ELres}).
For the Flat $\duzero$-distribution, the result is even simpler, as the bracketed expression in Eq.~(\ref{probnearzone})
is replaced by $\beta/\bu$ for this case.
The near zone expression for $\Delta n$ is
    \beq\label{Deltannearzone}
        \Delta n=  \min\left\{ 1, \frac{3\beta n^3}{4\pi\Nlam v^2} \right\}\,.
    \eeq
As discussed below Eq.~(\ref{Deltansmallv}), if coherent oscillations are sought, then $v\alt 1/20$
implies that $\min\{\cdots\}$ in Eq.~(\ref{Deltannearzone}) is unity, and so $\Delta n=1$ for all $n$.
\par
It is worth noting that the oscillation probability of Eq.~(\ref{probnearzone})
(and the mixing probability that results from this equation by setting
$\langle\sin\frac{L\delta\tilde H_n}{2}\rangle=0$,
$\langle\sin^2\frac{L\delta\tilde H_n}{2}\rangle=\frac{1}{2}$),
have an $n$-independent factor of $v^2$.
Using Eq.~(\ref{Eresnearzone}), this factor can be written as
\beq\label{vsq}
v^2 = \frac{3\Delta m^2 \cos 2\theta}{\Eres^2 (n=1)}\,.
\eeq
The meaning of this exercise is that the oscillation probability (and mixing probability) is proportional to
$\Delta m^2/\Eres^2 (n=1)$.  Thus, at least in the case of the Flat~$\duzero$-distribution for which there are no
other factors exceeding unity, the energy of the first resonance cannot be too much larger than the sterile
neutrino mass if the mixing is to be observable.  For example, for an oscillation probability of 0.3\%,
the energy of the first resonance must occur at about thirty times the sterile neutrino mass.

\subsection{Convergence of the sum of amplitudes}
\label{subsec:convergence}

It is well known that the Riemann sum $\zeta(s)\equiv\sum_{n=1}^\infty (1/n^s)$ is convergent
for $s>1$, and divergent for $s\le 1$.  In this subsection, we investigate the convergence properties
of $P_{as}$ as given in Eq.~(\ref{probnearzone}).
For simplicity, we will assume the more common outcome of Eq.~(\ref{Deltannearzone}) and set $\Delta n =1$.
Since $\Delta n$ is bounded from above by unity, this assumption makes little difference in
what follows.
Also, the two sine factors in Eq.~(\ref{probnearzone}) are bounded in magnitude by unity,
and so will not be included in what follows.
We are left to analyze the rates of convergence of
    \beq\label{convfactor}
        \Sigma (N) \equiv \sum_{n=1}^N \left( \frac{v}{n^2} \right)
        \left\{
        \begin{array}{lll}
        	   \sqrt{\frac{2}{\pi}}\,\frac{\beta E}{\sigma}
            ~ e^{-\frac{1}{2}\left(\frac{\beta E v}{\sigma n}\right)^2}\,,
        & {\rm Gau{\ss}ian\ }\duzero {\rm -distribution}\, \\
        & \\
           \frac{\beta}{\bu}\,, & {\rm Flat\ }\duzero {\rm -distribution}\,, \\
           \end{array}
	\right.
	\eeq
for our two $\duzero$-distributions.
The analysis is quite different for the Flat and Gau{\ss}ian distributions.

\subsubsection{Convergence of Flat $\duzero$-distribution}
\label{subsubsec:flat-u}

With the Flat $\duzero$-distribution,
it is clear that the summation is convergent, with each term in Eq.~(\ref{convfactor}) weighted by $1/n^2$.
The convergence of the sum is modulated somewhat by the two sine factors in Eq.~(\ref{probnearzone}).
\par
It is true that any experiment will have a resonant mode, at $n_R \sim (LE)/\ELres$, where
$\ELres$ is the constant defined in Eq.~(\ref{ELres}).
However, for the modes with $n\ll n_R$, the mixing factor $\sin 2\tilde\theta_n$ is very suppressed,
while for the modes with $n\gg n_R$, both sine factors approach the 4D~vacuum values.
Since the resonant amplitude is suppressed by $1/n_R^2$, the resonance feature will not
be observable unless $n_R$ is a small integer.
Thus, for the Flat $\duzero$-distribution case, the first few modes will dominate the sum,
and we may expect that the $n=1$ resonance at $LE =\ELres$ is the most manifest.

\subsubsection{Convergence of Gau{\ss}ian $\duzero$-distribution}
\label{subsubsec:Gaussian-u}

For the Gau{\ss}ian~$\duzero$-distribution, we may write
    \beq\label{SigmaGauss}
        \Sigma (N)=\sum_{n=1}^N \left(\frac{1}{n}\right) \sqrt{\frac{2}{\pi}}\,\left | \frac{d}{dz} \left(e^{-z^2/2}\right) \right |_{
        z=\left( \frac{\beta E kL}{2\sigma n} \right) }\,.
    \eeq
Here the rate of convergence of the series depends critically on the slope of the Gau{\ss}ian~$e^{-z^2/2}$.
The magnitude of the slope monotonically increases from zero at $z=0$ and $\infty$,
to its maximum value of $1/\sqrt{e}$ at $z=1$ (the inflection point~``IP'').
As $n$ increases, $z$ decreases in inverse proportion.
Thus, in the region $z>1$, increasing $n$ will drive the magnitude of the slope to larger values,
and the $(n+1)^{st}$ term in the series will be at least as large as $\frac{n}{(n+1)}$ times the previous term.
On the other hand,  for $z \le 1$, increasing $n$ will drive the magnitude of the slope to smaller values,
and the series will converge faster than $1/n$.
The consequence is that the series is mathematically convergent,
but all amplitudes with $z\agt 1$ will contribute significantly to the sum.
An equivalent statement is that all amplitudes up to
    \beq\label{nIP}
        n\sim \nIP = \frac{\beta LE k}{2\sigma} =
        1000\,\beta\,\left(\frac{LE}{0.4\;\rm km GeV}\right)
      	\left(\frac{k}{10^{-9}{\rm m}^{-1}}\right) \left(\frac{\Delta u}{\mu{\rm m}}\right)
    \eeq
contribute significantly to the sum.
One sees that, depending on the variables $LE,\ k$ and $1/\sigma\sim \Delta u$,
the number of significant amplitudes, $\nIP$, may be large.
Moreover, the number of significant amplitudes scales with $LE$.

\subsection{Near zone phenomenology}
\label{pheno}
\par
It is illuminating to present examples of the new phenomenon
arising from the $LE$-resonance of the warped extra-dimensional model.
What becomes important for each experiment is the product of baseline times neutrino energy.
In Table~(\ref{table:one}) we list some baseline and energy parameters for several recent and proposed experiments~\footnote
{
We do not consider solar neutrinos because the accepted theory of matter effects in the Sun leads to
emission of solar neutrinos which are nearly 100\% in the mass eigenstate $\nu_2$.
Mass eigenstates do not oscillate.
}.
One possible phenomenology would be a resonant value of $LE$ between the LSND and MiniBooNE values of
$2.5\times 10^{-3}$~km~GeV and $2.5\times 10^{-1}$~km~GeV, respectively.
With such a resonant value of $LE$,
active-sterile vacuum oscillation, or even the resonance, could explain the LSND excess,
while no observable active-sterile mixing would be expected in MiniBooNE, or in any other longer-$LE$ experiments.
In particular, active-sterile mixing would be suppressed in the SuperK data sample of atmospheric neutrinos.

We leave for future work the matching of the model with existing neutrino oscillation data.
Here we explore the $LE$~resonant phenomenon without reference to data.
As with the discussion above of convergence rates for amplitude sums,
the discussion of phenomenology bifurcates
for the two $\duzero$-distributions which we employ.

\begin{table}
\begin{tabular}{||l|c|c|c||c|c|c|c|c||}
\hline\hline
Experiment & Baseline\ $L~[{\rm m}]$ & Energy\ $E~[{\rm MeV}]$ & Product\ $LE~[{\rm km GeV}]$ & $\lamosc~[{\rm m}]$
	& $\Nlam=L/\lamosc$ & \\ \hline\hline
Bugey & 25  & 1$-$6 & $(2.5-15)\times 10^{-5} $  & 2.5$-$15 & 1.7$-$10 &  \\ \hline
KARMEN & 17.7  & 20$-$52.8 & (0.35$-$0.93)$\times 10^{-3} $  & 50$-$130 & 0.13$-$0.35 & \\ \hline
LSND & 30 & 20$-$50 & $(0.6-1.5)\times 10^{-3}$ & 50$-$125 & 0.24$-$0.6 &\\ \hline
SNSosc\ (ORNL) -- near & 18  & 15$-$53 & $(0.27-0.95)\times 10^{-3}$  & 38$-$130 & 0.14$-$0.47 &  \\ \hline
SNSosc -- far & 60  & 15$-$53 & $(0.90-3.2)\times 10^{-3}$  & 38$-$133 & 0.47$-$1.6 &  \\ \hline
FINeSSE\ (FNAL) -- near & 20  & 500$-$1000 & $(1-2)\times 10^{-2}$  & 1300$-$2500 & 0.008$-$0.015 &  \\ \hline
FINeSSE -- far & 100  & 500$-$1000 & 0.05$-$0.10  & 1300$-$2500 & 0.040$-$0.077 & \\ \hline
MiniBooNE &  540 & 300$-$800 & 0.16$-$0.43 & 740$-$2000 &  0.3$-$0.7 & \\ \hline
CDHS & 755  & $>$~1000 & $>$~0.76  & $>$~2500 & $<$~0.3 & \\ \hline
Kamland\ (dominant reactor) & $180\times 10^3$  & 1$-$7 & 0.18$-$1.26  & 2.5$-$17.6 & $(1-7)\times 10^4$ & \\ \hline
SuperK\ (subGeV atmos.) & $(10-10^4)\times 10^3$  & 500$-$1000 & 5.0$-10^4$ & 1200$-$2500
	& 4$-$8000 & \\ \hline
Minos & $735\times 10^3$  & $(1-50)\times 10^3$ & $730-3.7\times 10^4$  & $(2.5-125)\times 10^3$  & 6$-$300 & \\ \hline\hline
\end{tabular}
\caption{Baseline $L$, energy range $E$, product $LE$, 4D vacuum oscillation length $\lamosc$,
and the number $\Nlam$ of vacuum oscillation lengths in the baseline,
for several neutrino data sets~\cite{Armbruster:2002mp, Aguilar:2001ty, Garvey:2005pn, AguilarArevalo:2007it, :2008ee, Michael:2006rx, Declais:1994su, Abramowicz:1984yk, Bugel:2004yk, Ashie:2005ik}, in order of increasing $LE$.
For $\lamosc$ and $\Nlam$, a neutrino mass-squared difference of 1 eV$^2$ is assumed.
The SNSosc experiment is a neutrino detector proposed for the Spallation Neutron Source
at Oak Ridge National Laboratory.}
\label{table:one}
\end{table}

\subsubsection{Phenomenology with Flat $\duzero$-distribution}
\label{flatpheno}

\begin{figure}
\centering
\raisebox{4.8cm}{${P_{as}}$}
\includegraphics[scale=0.31]{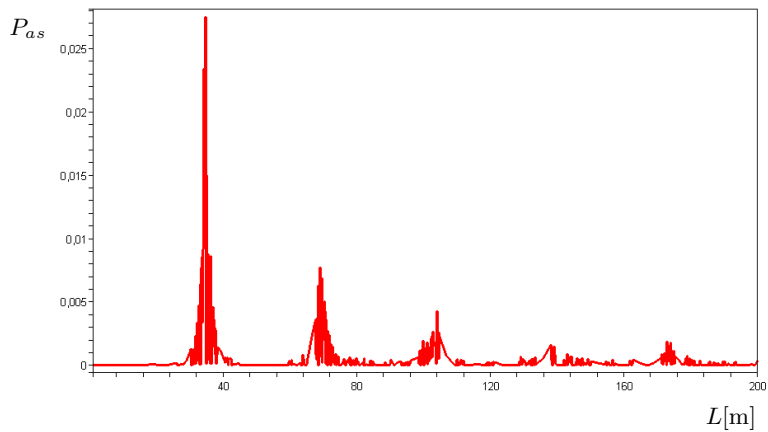}
\\{\hspace{9cm}${L[\text{m}]}$}
\caption{Oscillation probability as a function of the baseline $L$ for an assumed flat distribution of $\duzero$.  Parameter values are
$\bu=\beta=1$, $\text{sin}^2 2\theta=0.003$, $k=5/\text{km}$, $E=$~MeV, and $\Delta m^2=2500~\text{keV}^2$. The resulting value of $\ELres$ is 35~m~MeV.
For our choice of $E$, the resonance peaks occur at $L=n\,\ELres/E=35n$~m, $n=1,2,3\dots$,
with the principal resonance corresponding to $n=1$.}
\label{fig:flatn1}
\end{figure}

With the Flat $\duzero$-distribution, the $1/n^2$ falloff in amplitude implies that the small values of $n$ provide the dominant contributions to the oscillation probability.
In particular,  $n=1$ defines the most dominant ``principal resonance'', as can be seen in
Fig.~\ref{fig:flatn1}.
The peaks in the oscillation probability are observed at distances equal to integer multiples of $\ELres/E$.  The heights of the successive peaks decrease rapidly with increasing $n$,
roughly as $1/n^2$.  We have chosen parameters so that the peak of the principal resonance
is of order of the mixing parameter inferred from the LSND excess.
Attainment of this order of magnitude within the context of the new model is a demonstration of the potential relevance of the model to the real world.
\par
To the extent that the principal resonance is dominant,
the phenomenology is relatively straightforward.
At fixed baseline, the principal resonant mixing occurs at a unique energy, with 5D vacuum oscillations
pertaining at energies well below this principal resonant energy,
and suppression of mixing pertaining at energies well above.
This part is standard oscillation phenomenology~\footnote
{
But recall that, in contrast to the 4D~vacuum oscillation formula,
the 5D~vacuum oscillation formula includes the $\duzero$-distribution factor of $\int dN_D$
and the path-integral weighting $e^{iS_{cl}(n)}$.  These additional factors are discussed in
subsection~\ref{subsec:5Dlimit}.
}.
What is truly new is that at fixed neutrino energy, resonant mixing occurs at one baseline value $L$,
with vacuum oscillations pertaining at much smaller baselines and suppression of mixing pertaining for much larger baselines.
So the sterile neutrino decouples from active neutrino physics at either sufficiently large $E$ or sufficiently large $L$.
The principal resonance is not an energy resonance or a length resonance, but rather a product $LE$ resonance.
So a more careful statement is that once the resonant value of $LE$ has been identified experimentally,
then 5D~vacuum oscillations apply at smaller $LE$, and active-sterile mixing is suppressed at larger $LE$.
\par
One consequence of the principal resonance condition $LE\approx \ELres$ is that,
should an experiment observe this resonance, then in the context of this model one may infer the value of the
warp-parameter.  From  Eq.~(\ref{ELres}), we have that $k=\sqrt{12\Delta m^2 \cos 2\theta}/\ELres$.
Corroborative information may be carried by the height of the principal resonance peak.
From Eq.~(\ref{ELres}), one easily derives that on the principal resonance,
$(vE)_{\rm Res}=\sqrt{3\Delta m^2 \cos 2\theta}$, and so
    \beq\label{firstpeak}
        (P_{as})_{\rm first\ peak} = [ \Delta n (n=1) ]^2
        \left[ \left(\frac{kL}{2}\right)^2 = \ \frac{3\Delta m^2 \cos 2\theta}{\Eres^2 (n=1)} \right]
	      \sin^2 \left(\frac{L\Delta m^2 \sin 2\theta}{4E}\right)
    \eeq
The $L$ and $E$ appearing in this formula
are correlated by the resonance condition Eq.~(\ref{ELres}).

\subsubsection{Phenomenology with Gau{\ss}ian $\duzero$-distribution}
\label{Gaussianpheno}
\begin{figure}
\centering
\raisebox{5.2cm}{${P_{as}}$}
\includegraphics[scale=0.35]{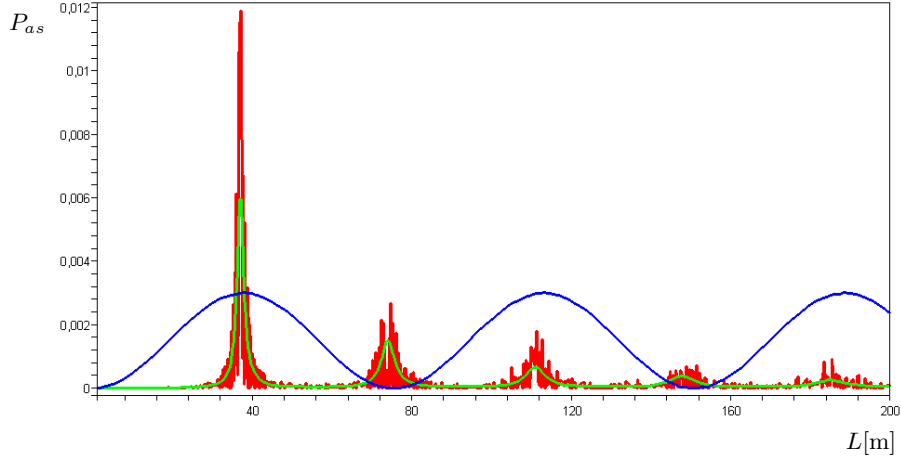}
\\{\hspace{11cm}${L[\text{m}]}$}
\caption{Oscillation probability as a function of the experimental baseline, for the Gau{\ss}ian distribution (red and green curves). The green curve presents the phase-averaged oscillation probability, and the sinusoidal blue curve presents the probability as given by the standard 4D~vacuum formula for oscillations between sterile and active neutrinos. Parameter choices are $\text{sin}^2 2\theta=0.003$, $k=5/(10^{8}~\text{m})$, $E=15$~MeV, $\Delta m^2=64 ~\text{eV}^2$,
and $\sigma=100~\text{eV}$. The resulting value of $\ELres$ is 550~m~MeV.
For our choice of $E$, the resonance peaks are found at the multiples  $L=n\ELres/E=37n$~m,
$n=1,2,3\cdots$, with the principal resonance corresponding to $n=1$.} \label{fig:gaussian1}
\end{figure}

The oscillation probability versus baseline for a Gau{\ss}ian~$\duzero$-distribution is
shown in Fig.~\ref{fig:gaussian1} (red curve).
With the Gau{\ss}ian~$\duzero$-distribution, there may be many modes making significant
contributions to the overall amplitude, in which case an analytical analysis seems difficult.
The number of significant modes is indicated in Eq.~(\ref{nIP}).
For the parameters we use in Fig.~\ref{fig:gaussian1}, the number of significant modes is of the order of
$n_{IP}=27\,(L/200{\rm m})$.  For the principal resonance at $L=37$~m, $n_{IP}\sim 5$,
whereas for the fifth resonance at $L=5\times 37$~m, one gets $n_{IP}\sim 25$.\\
Still,  Fig.~\ref{fig:gaussian1} reveals an oscillation probability with peak heights and peak structures not dissimilar to the case of the flat distribution (although parameter choices are quite different in the two cases).
The peaks are again found at multiple integers of the principal resonant product $\ELres/E$.
\par
Shown in green in Fig.~\ref{fig:gaussian1} is the phase-averaged oscillation probability.  For our parameter choices, the phase-averaged plot provides a faithful representation of the oscillation probability even for the low-energy principal resonance.
\par
We also show (in blue) in Fig.~\ref{fig:gaussian1}
the 4D~vacuum model of active-sterile oscillation, given in Eq.~(\ref{4Dvacuumprob}).
It is well-known that the two-state active-sterile 4D~vacuum solution,
while it can explain the LSND data in isolation,
cannot explain the comprehensive sets of data, including null results of other short-baseline
experiments~\cite{no4D}.
Still, we show it for comparison to our extra-dimensional model.
Large differences between the two models are apparent.
In particular, the 4D~vacuum result has maxima where the argument of $\sin^2 (L\Delta m^2/4E)$
equals odd multiples of $\pi/2$, i.e., at $L=(2\pi E/\Delta m^2)\times q$, with $q=1,3,5,\dots$.
On the other hand, the extra-dimensional model has maxima roughly where $\sin\tilde\theta$ equals unity,
i.e., at $LE=\ELres\times n$, with $n=1,2,3,\dots$.
Hence, within a fixed baseline or energy range, there are on average twice as many peaks in the extra-dimensional
model as in the 4D~vacuum model.
\par
\begin{figure}
\centering
\includegraphics[scale=0.35]{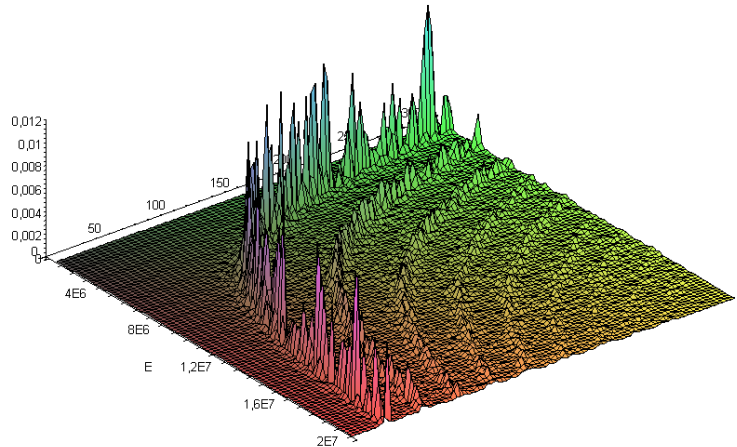}
\caption{Oscillation probability (vertical) in the $L$-$E$~plane, with the same parameters as in Fig.~\ref{fig:gaussian1}.
Units of $L$ and $E$ are m and eV, respectively.} \label{fig:gaussian_distribution3D}
\end{figure}

The novelty of the $LE$ resonance is best illustrated in a three-dimensional plot, with oscillation probabilities rising as topography on a two-dimensional $L$-$E$~plane.
Contours of constant $LE$ appear on this plot as hyperbolas symmetric about the diagonal
line $L=E$.  With the Gau{\ss}ian $\duzero$-distribution, inspection of the near-zone formula in Eq.~(\ref{probnearzone})
reveals that scaling in $LE$ (recall that $v=kL/2$) is broken only by the explicit $L$ in the argument of
$\sin(L\delta\tilde H_n/2)$ (or equivalently, by an explicit $E$ if we write $L=(LE)/E$).  Thus, we expect
the oscillation probability to roughly peak at resonant values of $LE$, i.e., along hyperbolic contours of
constant $LE$.  Such peaking is clearly evident in  Fig.~\ref{fig:gaussian_distribution3D}.
The hyperbolic contours of resonant probability display a marked contrast from, say, MSW resonances
which occur at resonant energy values independent of baseline $L$;
and the hyperbolic iso-$LE$ contours contrast with 4D~vacuum maxima,
which occur at $\sin(L\Delta m^2/4E)=1$,
i.e., at rays in the $L$-$E$~plane given by $L/E=q\,2\pi/\Delta m^2$, $q=1,3,5\dots$

We end this section with a comment on the possible slow convergence of the mode sum in the case of the Gau{\ss}ian
$\duzero$-distribution.
If the parameters are such that $n_{IP}\equiv \beta LEk/2\sigma$ defined in Eq.~(\ref{nIP}) is large compared to one,
then there are several or more contributing modes, and the concept of principal resonance gives way to
a contributing band of resonances about the mode number
    \beq\label{resband}
        \nRband\equiv \frac{LE}{\ELres}\,,
    \eeq
with $\ELres$ given in Eq.~(\ref{ELres}).
For the resonances in this band, the amplitudes are enhanced compared to their 4D vacuum analog by the factor
    \beq\label{bandenhance}
        \frac{\sin 2\tilde\theta_{\nRband} \,\sin \frac{L\delta{\tilde H}_{\nRband}}{2}}
 	      {\sin 2\theta \,\sin \frac{L\delta H}{2}}
        \simeq
        \frac{\sin \left(\pi\Nlam\sin 2\theta\right)} {\sin 2\theta \,\sin \left(\pi \Nlam \right)} \,.
    \eeq
For small $\theta$, but large $\pi\Nlam\sin 2\theta \agt 1$, this enhancement of the amplitudes in the resonant band
can be quite large, typically of order $1/2\theta$.  This enhancement is presumably operative in Figs.~\ref{fig:gaussian1}
and \ref{fig:gaussian_distribution3D}, where, as we have stated above, $n_{IP}$ ranges from 5 at the first resonance band
to 25 at the fifth resonance band.

\section{Discussion and conclusion}
\label{sec:discussion}
The basic feature of the new model presented herein
is that longer neutrino travel times associated with  longer baselines on the brane
allow the off-brane geodesic of the sterile neutrino to plunge deeper into the bulk and experience a greater warp factor.
As a consequence, the shortcut parameter $\epsilon=\delta t/t$ for the sterile neutrino increases with the baseline,
corresponding in turn to a decreasing resonance energy in the effective Hamiltonian of the two-neutrino system.
In addition, there are higher energy/longer baseline resonances resulting from additional classical geodesics.
We solved the geodesic equations and identified the countably infinite number geodesics
giving rise to the countably infinite
number of resonances.
Then we performed a path-integral-weighted sum over the amplitudes resulting from each of the geodesics.

In the ``near zone'', defined as baselines short on the scale of the warp factor $k^{-1}$, the resonance condition is
that the product of baseline and energy, $LE$, be an integer multiple of a fundamental value $\ELres\sim m_s/k$.
That the brane-bulk resonances in the "near zone" expansion depend on the product of the energy and baseline,
rather than on the energy alone as with the MSW matter-resonance, is a novel feature of our model.

Whether or not the higher resonances coming from the additional geodesics contribute significantly depends on
the initial distribution of sterile velocities $\duzero$ transverse to the brane.
We considered two very different distributions for initial $\duzero$,
a Gau{\ss}ian distribution with a width related via the uncertainty principle to the brane width,
and a Flat distribution as might arise in a path-integral approach.
Both approaches led to qualitatively similar oscillation probabilities, although the parameters in the two approaches,
and the physical interpretation of some of the parameters, were quite different.
Also, the contributions of higher resonances are more suppressed in general for the Flat distribution
than for the Gau{\ss}ian distribution.
\par
Since higher-$LE$ resonances are suppressed, and active-sterile neutrino mixing is suppressed for $LE$
above the resonant values, sterile neutrinos taking shortcuts in the extra-dimensional bulk
decouple from active neutrinos in long-baseline experiments.
Thus, no active-sterile mixing is expected in atmospheric data, in MINOS or CDHS.
All explanations proposed  so-far for the LSND and MiniBooNE anomalies assumed baseline-independent
oscillations and mixing.  Difficulties accommodating longer-baseline data were encountered in these models.
These difficulties do not immediately extend to scenarios with warped extra dimensions, as developed here.
In fact, the failure of previous models to reconcile short baseline data such as LSND with
longer baseline data might be construed as favoring the extra-dimensional shortcut scenario.
Finally, we mention that the bulk shortcut scenario might even relieve some of the remaining tension between
the LSND and KARMEN experiments.
Since LSND has almost twice the baseline of KARMEN, the bulk shortcut model opens more parameter space for accommodating the two experiments.
\par
In conclusion, we have found that in scenarios with sterile neutrinos taking
shortcuts in the extra-dimensional bulk, the shortcut is
baseline $L$ dependent as well as energy $E$ dependent.
Resonances occur as a function of $L$ as well $E$, and may lead to neutrino anomalies.
Above the contributing resonances, i.e., at large $L$ or large $E$,
the mixing sterile neutrinos with active neutrinos is suppressed.
In the context of this model, existing data on neutrino oscillations and anomalies need to be reanalyzed.
We intend to compare the model to the world's data in future work.

\acknowledgments
During the course of this work,
TJW was supported by the US Department of Energy grant DE-FG05-85ER40226,
an Alexander von Humboldt Foundation Senior Research Award,
the faculty leave program of Vanderbilt University,
and the hospitality of the Technische Universit\"at Dortmund and the Max-Planck-Institut f\"ur Physik
(Werner-Heisenberg-Institut), M\"unchen, and f\"ur Kernphysik, Heidelberg.
We thank John G. Learned and Danny Marfatia for useful comments in the very early stage of this work.

\end{document}